\documentclass{elsart}
\begin{document}
\input epsf
\begin{frontmatter}
\title{Synchronization in large populations of limit cycle oscillators with long-range interactions}
\author{M.S.O. Massunaga}
\address{Laborat\'orio de Ci\^encias F\'\i sicas\\
 Universidade
Estadual do Norte Fluminense\\
Av. Alberto Lamego 2000, Campos dos Goytacazes, RJ, Brazil, 28015-620}
\author{M. Bahiana}
\address{Instituto de F\'\i sica\\
 Universidade Federal do Rio de Janeiro\\
 Caixa Postal 68528, Rio de Janeiro, RJ, Brazil, 21945-970}
\date{\today}
\begin{abstract}
We study the onset of synchronization in lattices of limit cycle oscillators with long-range coupling by means of numerical simulations. In this regime the critical coupling strength depends on the system size and interaction range reflecting the non extensive behavior of the system, but an adequate scaling removes the dependency and collapses the long-range synchronization curves with the one resulting from a system with uniform coupling. 
Two descriptions are considered, the Kuramoto or phase description model and a Cell Dynamical System model for phases and amplitudes. Oscillator death is observed in the second approach.
\end{abstract}
%
%
\end{frontmatter}
\section{Introduction}
Coupled limit-cycle oscillators have been extensively used to model the collective behavior of systems in physics \cite{fisher}, chemistry and biological sciences \cite{glass-book1,win-book1}.
The existence of a self organized periodic motion is a direct consequence of the energy exchange between the system and its surroundings keeping the system far from equilibrium. The mathematical description of those systems relies on differential equations, or its discrete versions, since no Hamiltonian exists, and the appearance of a macroscopically synchronized state has been treated as non-equilibrium phase transition \cite{win-book1,kura-book}. The coupling among oscillatory units is usually considered to be uniform and of infinite range or with first neighbors only. The reason for these choices is simplicity, uniform coupling makes possible  analytical calculations, and nearest neighbor coupling is a computationally efficient form of short range coupling. Here we would like to consider a long range distance dependent coupling of the form $r^{-\alpha}$, where $r$ is the distance between oscillators and $\alpha>0$ defines the range: $\alpha=0$ corresponds to uniform coupling, and $\alpha\rightarrow\infty$ to nearest neighbor coupling. For this we consider two different approaches: the Kuramoto model for the phases and a cell dynamical system (CDS) model for phases and amplitudes along the limit cycle. We observe that, as in equilibrium systems, the critical value for synchronization will depend on the system size and on $\alpha$. For both models we show that a scaling proposed for equilibrium nonextensive spin systems works well for the synchronization process. 
\section{Long Range Coupling in Equilibrium Spin Systems}
It well known that, depending on the dimensionality $d$ of the system, interacting potentials of the type $J/r^{-\alpha}$ may lead to nonextensive energies. This problem is usually avoided in the case of uniform or mean field coupling, for which $\alpha=0$, by the redefinition of the coupling term with the division by the size of the system, $N$. This correction may be extended to other values of $\alpha$ in the nonextensive regime with the definition of a $N^*(\alpha,d)$ that may be calculated from the requirement that the total energy of the system must be finite.
The equilibrium properties of chains of ferromagnetically coupled spins have been considered in the nonextensive regime for the Ising model \cite{nobre-tsallis,cannas-tamarit} and XY model \cite{tamarit-ante,campa}, the finiteness condition for the energy leads to a common correction factor 
\begin{equation}
N^*(\alpha,N)=\int ^{N^{1/d}}_1 dr r^{d-1}r^{-\alpha}=\label{scaling}
\frac{N^{1-\alpha/d}-1}{1-\alpha/d}.
\end{equation}
The extensive behavior is recovered with the definition of the scaled interaction strength $J'=J/(2^\alpha N^*)$.
For $\alpha>d$, (\ref{scaling}) provides a correction for finite size effects, and for $\alpha<d$ it removes the dependency of the critical coupling strength on $\alpha$ and $N$. In the limit $N\rightarrow\infty$, $N^*=N$ for mean-field coupling ($\alpha=0$) and, $N^*=\ln N$ for $\alpha=d$. The behavior of thermodynamical quantities like magnetization and energy as a function of temperature $T$ for different values of $N$ and $\alpha$ are found to collapse with the mean field curve when plotted as a function of $T/(N^*2^\alpha)$, indicating that this regime contains all the information regarding the nonextensive behavior.
\section{Kuramoto Model}
One of the simplest models for systems of coupled limit-cycle oscillators is the Kuramoto model for the time evolution of the phases. As shown in \cite{kura-book}, it possible to reduce any system of $N$ weakly coupled limit cycle oscillators near the Hopf bifurcation, to the set of equations
\begin{equation}
\dot{\theta}_i=\omega_i+\sum_{j}^{N}\Gamma_{ij}(\theta_j-\theta_i),\;\;\;(i=1,2,\ldots,N)\label{kmodel}
\end{equation}
where $\theta_i$ and $\omega_i$ are the phase and natural frequency of the $ith$ oscillator respectively. The coupling term $\Gamma_{ij}(\theta_i-\theta_j)$ is a periodic function that may be obtained from the original equations of motion, regardless of the dimensionality of the oscillating field. The case of uniform, or mean field,  coupling has been considered with detail by Kuramoto \cite{kura-book} for the simplest form of the coupling term,
\begin{equation}
\Gamma_{ij}=N^{-1}K\sin(\theta_j-\theta_i). \;\;\; (K>0)\label{coupling}
\end{equation}
Kuramoto considered the possibility of a collective oscillatory motion when the natural frequencies are picked from a given distribution $g(\omega)$, as would happen in a real system. As the coupling modifies the frequencies, acting as an external oscillatory force, it is necessary to define the asymptotic frequency of the $ith$ oscillator after the transient regime as
\begin{equation}
\tilde{\omega}_i=\lim_{T\rightarrow\infty}\frac{1}{T}\left[\theta_i(t+T)-\theta_i(t)\right].\label{afreq}
\end{equation}
The existence of a collective behavior with phase and frequency entrainment depends on the coupling strength and on the choice of $g(\omega)$ \cite{kura-book,dai1}. In what follows we consider only the case where $g(\omega)$ is Gaussian.
 Kuramoto characterized the degree of synchronization in the asymptotic regime  by means of two order parameter like quantities: $R$ for the frequencies, and $\sigma$ for the phases. Frequency entrainment may be quantified simply by analyzing the formation of frequency clusters  and defining
\begin{equation}
R=\lim_{N\rightarrow \infty}\frac{N_s}{N},
\end{equation}
where $N_s$ is the size of the largest cluster of oscillators with a common asymptotic frequency $\tilde{\omega}$. 
The complex quantity
\begin{equation}
\sigma =\frac{1}{N}\sum^N_{j=1}\exp(i\theta_j)
\end{equation}
is appropriate for the phase ordering as it measures the concentration of phases on a certain value.
$R$ and $\sigma$ behave as  equilibrium order parameters in the sense that, for $K$ below some critical value $K_c$, oscillations are incoherent leading to $R$ and $\sigma$=0. For strong enough coupling, or $K>K_c$, collective oscillation takes place and $R,\sigma>0$. 

For the case of uniform coupling, phase and frequency entrainment appear simultaneously at the critical value
\begin{equation}
K_c=\frac{2}{\pi g(\omega_0)},\label{kc}
\end{equation}
where $\omega_0$ is the average value of the distribution of natural frequencies \cite{kura-book}.

The long range version of the coupling term is written as
\begin{equation}
\Gamma_{ij}=\frac{K}{r^\alpha_{ij}}\sin(\theta_j-\theta_i),\;\;\;K>0(r_{ij}=1,2,3\ldots),\;\; \label{necoupling}
\end{equation}
and was considered in the case of oscillator chains in \cite{rogers}.
In Hamiltonian systems the limit of non-extensivity can be found by requiring the upper value of the energy to be finite as $N\rightarrow \infty$. Limit cycle oscillators are open systems and energy is a meaningless quantity so, instead, we establish the finiteness condition for $(1/N)\sum_i\dot{\theta}_i$ in the asymptotic regime. This condition may be established by requiring that the upper value of the coupling term divided by $N$ must be finite, as in the mean field coupling case.
With this, we end up with a condition identical to (\ref{scaling}), but, as will be clear from the graphs below, the factor $2^{\alpha}$ does not seem to be present in this case.
For any value of $d$ and $\alpha$, it is then possible to incorporate the dependence on $\alpha$ and $N$ of the coupling term into the value of $K$  with the definition of a scaled coupling strength
\begin{equation}
K^*(\alpha,N)=K N^*. 
\end{equation}

In order to numerically study the non-extensive region, we used two dimensional square lattices of coupled oscillators with natural frequencies sorted from the Gaussian distribution  $g(\omega)=(1/\sqrt{2\pi})\exp(-\omega^2/2)$ and periodic boundary conditions. Equation (\ref{kmodel}) with the coupling (\ref{necoupling}) was then integrated using the Euler method with d$t=0.1$. The first 3000 iterations where discarded as a transient, and after 6000 iterations past the transient, we calculated the asymptotic frequencies as defined in (\ref{afreq}). We used the Hoshen-Kopelman method \cite{cluster} to analyze the frequency pattern, two sites were considered to belong to same cluster if the frequency difference between them was less than $2\pi/T$, where $T$ is the time elapsed after the transient. $R$ was then associated to the largest cluster, and averaged over 10 initial conditions for the natural frequencies. 

\begin{figure}[htb]
\leavevmode
\centering
\epsfxsize = 14.4cm
\epsfysize = 8.4cm
\epsffile{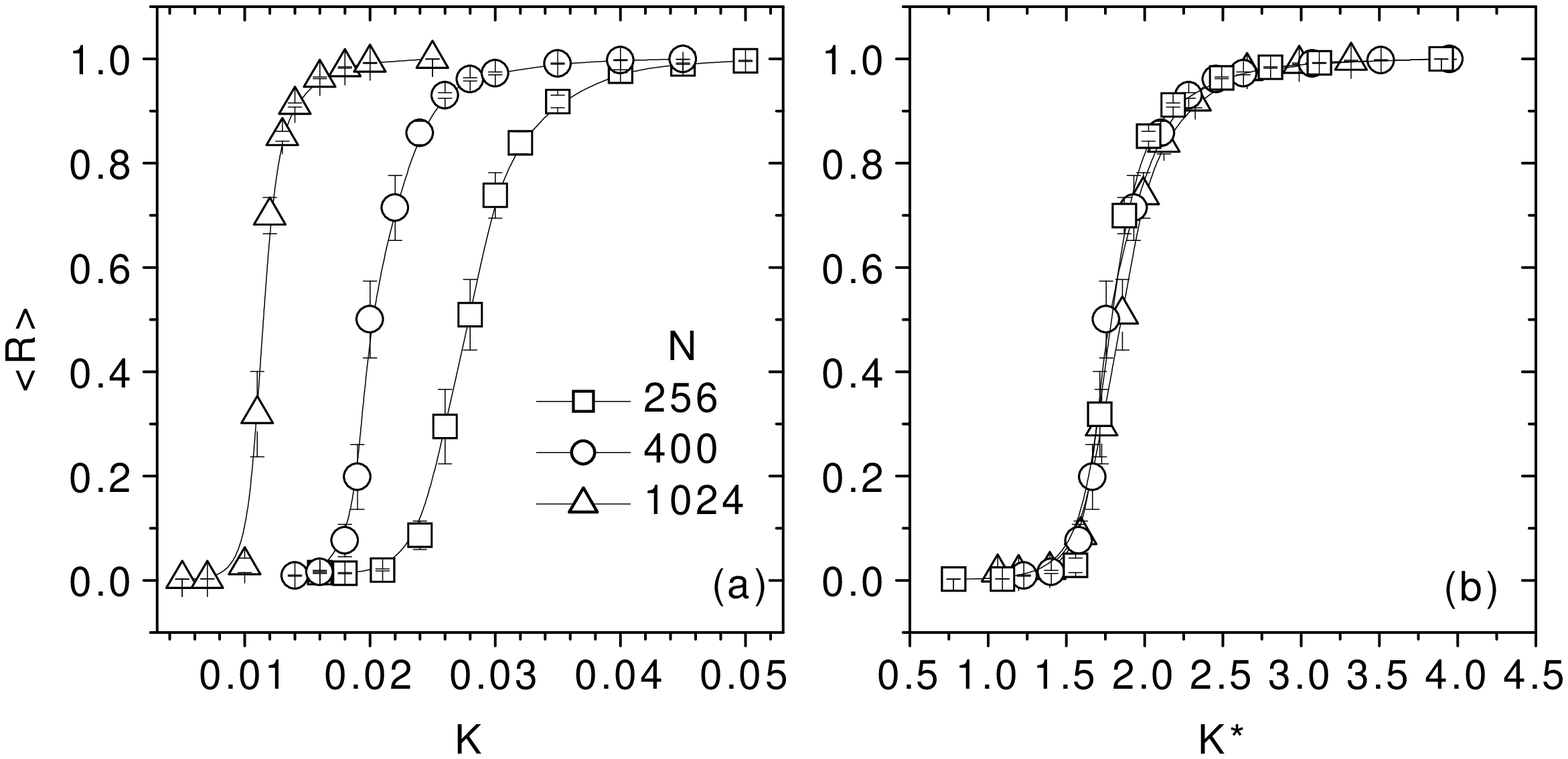}
\centering
\vspace{-2.5cm}
\caption{Frequency entrainment in square lattices with $N$ Kuramoto oscillators interacting through a long range coupling $K/r^{0.75}$. (a) The non-extensivity appears in the $N$ dependency of $K_c$, the critical value for synchronization. (b) The corrected coupling intensity, $K^*$, absorbs the non-extensive part of the coupling term leading to a synchronization curves independent of the system size.}
\label{alfa075}
\end{figure}

The non-extensivity is evident if we consider the behavior of $\langle R\rangle$ as a function of $K$ for $\alpha<d$. For fixed $\alpha$, the critical  value of $K_c$  decreases with increasing $N$ as can be seen in Fig. (\ref{alfa075}.a), as more terms contribute to the coupling. The shift in $K_c$ may be corrected with the use of (\ref{scaling}).
In fact, if we plot $\langle R\rangle$ as a function of $K^*$, there is no dependence on the system size, the curves collapse and a unique value of $K_c^*$ appears. Figure (\ref{alfa075}.b) shows the behavior of $\langle R\rangle$ as a function of $K^*$ for fixed $\alpha$.   
%
\begin{figure}[htb]
\leavevmode
\centering
\epsfxsize = 14.4cm
\epsfysize = 8.4cm
\epsffile{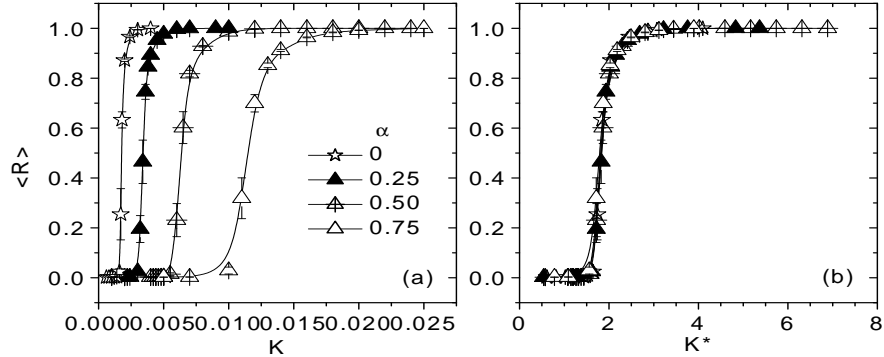}
\centering
\vspace{-2.5cm}
\caption{Frequency entrainment in square lattices with 1024 Kuramoto oscillators interacting through a long range coupling $K/r^{\alpha}$. (a) The non-extensivity appears in the $\alpha$ dependency of $K_c$, the critical value for synchronization. (b) The corrected coupling intensity, $K^*$, absorbs the non-extensive part of the coupling term leading to synchronization curves independent of $\alpha$ The curve for $\alpha=0$ correspond to uniform or mean-field coupling.}
\label{N1024}
\end{figure}
The dependence of $K_c$ on $\alpha$ is evident if we plot $\langle R\rangle$ for fixed $N$ and different values of $\alpha$. Figure (\ref{N1024}.a) shows the curves for $\alpha=0.25$, 0.50, 0.75, and 0 (mean-field coupling without the correction factor $1/N$). It is important to notice that we did not use the factor $2^\alpha$ as predicted for the spin systems. Consistently with dependence on system size, as $\alpha$ decreases, more terms are included in the coupling, leading to smaller values of $K_c$. Again, (\ref{scaling}) fixes the dependence on $\alpha$ providing a perfect collapse even with the mean-field curve as can be seen in Figure (\ref{N1024}.b). Notice that the value of $K^*_c$ in Figures (\ref{alfa075}) and (\ref{N1024}) is in good agreement with (\ref{kc}), our frequency distribution leads to $K^*_c= 1.59$.  
\section{Cell Dynamical System Model}
A more complete description of the system includes information about the amplitudes by means of the definition of an  
oscillatory field 
\[z(n,t)=\rho(n,t)\exp[i\phi(n,t)].
\] 
This is an important aspect since oscillator death is possible as we effectively increase the coupling strength with the variation of system size and coupling range \cite{erme1}. 

The spirit of cell dynamical modeling is the discretization of time and space from the start,  as opposed to formulating a continuous model and then further adopting some discretization scheme. In this sense, $(n,t)$ defines the position of the $n$-th cell at time $t$, $\rho$ and $\phi$ are then the amplitude and the phase of the field at given cell. The time evolution of the system is then given by the repeated application of the map
\begin{equation}
z(i,t+1)=F[z(i,t)]+\sum_{j\neq i}D_{ij}\{F[z(j,t)]-F[z(i,t)]\}
\label{cds}
\end{equation}
where the local dynamics is regulated by
$F[z(t);A]=A\exp(i \omega)\frac{z(t)}{\sqrt{1+|z(t)|^2(A^2-1)}}$. This form of $F$ corresponds to an asymptotically stable limit cycle of unit radius for all $A>1$ \cite{bamass2}. Here the parameter $A$ is a measure of the attraction to the limit cycle.
The long range coupling is implemented through the definition of  $D_{ij}=D_0/r_{ij}^\alpha$. 

\begin{figure}[htb]
\leavevmode
\centering
\epsfxsize = 14.4cm
\epsfysize = 10cm
\epsffile{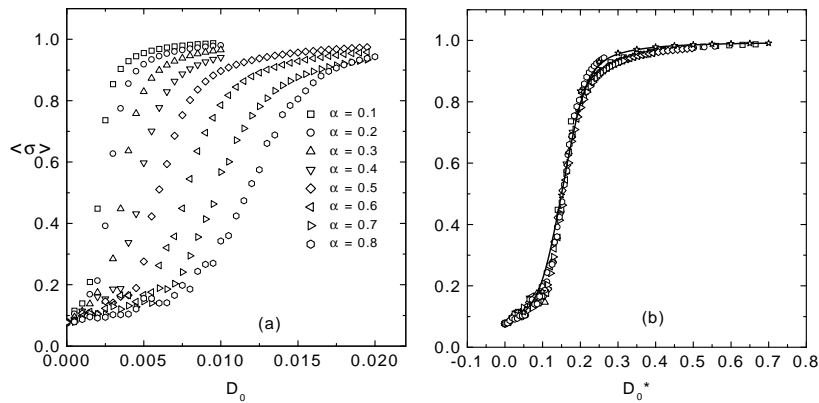}
\centering
\vspace{-3.0cm}
\caption{Phase entrainment in a CDS chain with  oscillators interacting through a long range coupling $D_0/r^{\alpha}$. (a) Smaller values of $\alpha$ lead smaller critical values of the critical coupling constant. (b) Again, the definition of the effective coupling constant, taking into account the nonextensive part of the coupling term provides the collapse of synchronization curves for different exponents.}
\label{chain100}
\end{figure}

For observation of synchronization, we considered chains of $N=100$ oscillators with periodic boundary conditions, $A = 2.0$, Gaussian distribution of natural frequencies with unit mean value and variance  0.1. We iterated (\ref{cds}) 4000 times,  the first  1000 being discarded as a transient, and  the order parameter $\sigma$ was then calculated. This procedure was repeated for different chain sizes and values of the decay exponent $\alpha$ for several values of the coupling strength $D_0$. Figure \ref{chain100}(a) shows the synchronization curves averaged over 20 samples. As expected, the critical value of $D_0$ is dependent on $N$ and $\alpha$.
Although the calculation of the upper bound of the coupling term is rather different from the one in the Kuramoto model or spin systems, the same form of scaling is observed here. Figure \ref{chain100}(b) shows the synchronization curves of $\sigma$ as a function of $D^*_0=D_02^\alpha N^*$.

\begin{figure}[htb]
\leavevmode
\centering
\epsfxsize = 13.cm
\epsfysize = 9.5cm
\epsffile{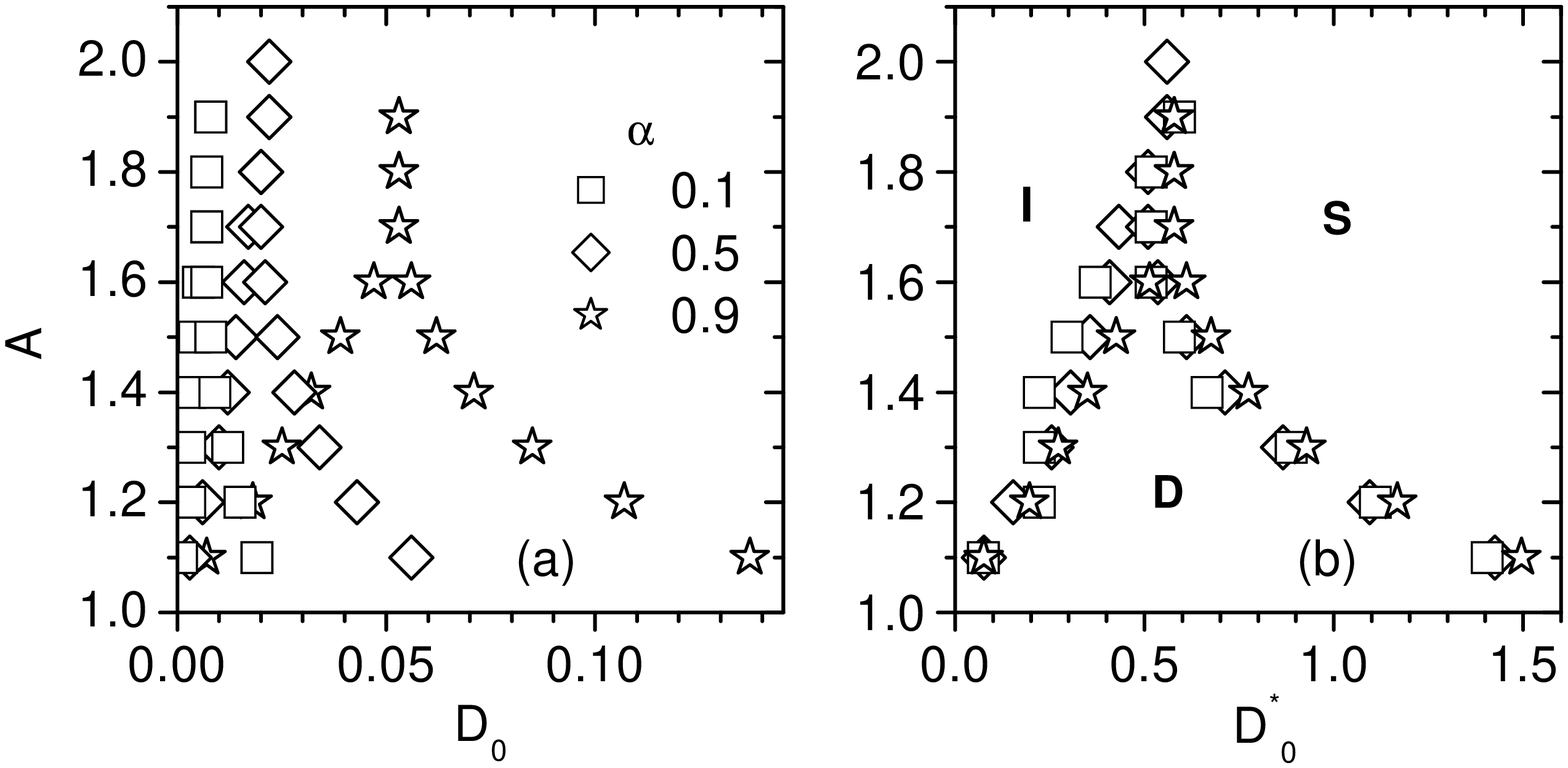}
\centering
\vspace{-2.7cm}
\caption{Phase diagram showing regions of synchronization, incoherent motion and oscillator death for a CDS chain  with  oscillators interacting through a long range coupling $D_0/r^{\alpha}$. Relative to each curve, right side region corresponds synchronized (S) motion, on the left side region oscillations are incoherent (I), and inside the lambda oscillators are dead (D).  (a) Smaller values of $\alpha$ corresponds to a shift to lower values of $D_0$ in the region of death. (b) The same phase diagram, now with the scaled coupling constant, showing the collapse of the diagrams for different exponents. }
\label{death}
\end{figure}

It is important to verify the effect of nonextensivity on the stability of the oscillatory state. It is well known that, depending on the amplitudes and frequency spread, an increase in coupling strength may lead to stabilization of the rest state. This has been observed for systems with mean field coupling \cite{erme1} and for pairs of oscillators \cite{bamass2,bar-eli}. In order to
observe this effect we increased the frequency spread to 0.7, then, for a fixed value of $N$ and $\alpha$, we varied $D_0$ and $A$. After the 4000 iterations we verified whether the amplitude was non zero. The same procedure was repeated for $\alpha=0.1$, 0.5 and 0.9. Figure \ref{death}(a) shows the $A-D_0$ phase diagram indicating regions of synchronization, incoherent motion and death for the different values of $\alpha$. The scaled phase diagram can be seen in  Fig. \ref{death}(b). 

\section{Conclusion}
The above results bring the process of synchronization even closer to an order disorder transition, as we directly use the scaling proposed for nonextensive equilibrium systems.
Also, we see that the important quantity for synchronization and stability of the oscillatory phase is the scaled coupling constant. With this in mind, one  can easily tune the interaction strength to obtain the desired regime, in the case of distant dependent coupling. Some points still remain unclear. First, there is no a priori reason for the scaling  of the CDS model to be the same as that of Hamiltonian spin systems. Second we cannot justify the absence of the factor $2^{\alpha}$ in the scaling of the Kuramoto model, which has a condition for extensivity identical to Hamiltonian spin systems. Actually we found by tentative and error that for the Kuramoto oscillators a better form of scaling should have a factor $2^{\alpha/d}$. Since the results for spin Hamiltonians  and for the CDS model are for one dimensional systems, this correction would 
not affect their scaling, although we are not able to analytically justify this form.

\section*{acknowledgments}
The authors thank Francisco Tamarit and Celia Anteneodo for interesting
discussions and
the Brazilian agencies CNPq,
FAPERJ and FUJB-UFRJ 
for financial support. 


\end{document}